\documentclass{acm_proc_article-sp}
\usepackage{amssymb,amsmath}
\usepackage{graphicx}
\usepackage{booktabs}
\usepackage{balance}
\usepackage{appendix}
\usepackage{color}
\usepackage{url}
\newcommand{\eat}[1]{}

\begin{document}

\title{Online Dating Recommendations: Matching Markets and Learning Preferences}

\numberofauthors{5} 
%
\author{%
K.\ Tu$^{*}$\mbox{\hspace{0.3cm}}
B.\ Ribeiro$^{\dag}$\mbox{\hspace{0.3cm}}
H.\ Jiang$^{\ddag}$\mbox{\hspace{0.3cm}}
X.\ Wang$^{\ddag}$\mbox{\hspace{0.3cm}}
D.\ Jensen$^{*}$\mbox{\hspace{0.3cm}}
B.\ Liu$^{\mathsection}$\mbox{\hspace{0.3cm}}
D.\ Towsley$^{*}$\\ \\
\begin{tabular}{cccccccc}
 $^*$UMASS &&
 $^\dag$CMU &&
 $^\ddag$Baihe.com  &&
 $^\mathsection$UML
\end{tabular}
}

\renewcommand\footnotemark{}
\renewcommand\footnoterule{}
\maketitle
\begin{abstract}
Recommendation systems for online dating have recently attracted much attention from the research community. 
In this paper we proposed a two-side matching framework for online dating recommendations and design an LDA model to learn the user preferences from the observed user messaging behavior and user profile features. 
Experimental results using data from a large online dating website shows that two-sided matching improves significantly the rate of successful matches by as much as 45\%. 
Finally, using simulated matchings we show that the the LDA model can correctly capture user preferences.
\end{abstract}


\section{Introduction}
Recommending a partner in an online dating website is a serious task.
Dating recommendations are fundamentally different from product recommendations.
For instance, in the extreme scenario where a TV celebrity decides to join a dating website, thousands of (male or female) suitors\footnote{We use suitor in a gender-neutral sense to define either male or female suitors.} would be interested in dating the celebrity.
But recommending the celebrity to thousands of suitors would be a recipe for disaster.
On one hand, the celebrity would be inundated with messages from suitors that he or she considers bad matches.
On the other hand, the rejected suitors would get frustrated to see their messages go unreplied.


The above anecdotal example exposes a deeper general challenge: to jointly match the expectations of both sides of this dating matching market\footnote{A precise definition of a matching market is given in Section~\ref{sec:background}.}.
Unfortunately, while the online dating literature has acknowledged the importance of receiver preferences (e.g.,~\cite{Alsaleh:2011hf,brozovsky2007recommender,Diaz:2010hf,krzywicki2010interaction,Nayak:hm,Pizzato:2010jg}), little progress has been made to learn these preferences from the data rather than relying on self-declared preferences which can be inaccurate~\cite{Slater13Love}.

In this work we put forth a probabilistic two-side dating market framework that,  through learned user preferences, is able to increase the chances of making successful matches.
In our framework we introduce an LDA probabilistic model of user preferences trained by the message exchanges between users.
This probabilistic model learns user preferences both through the general user features and the observed user-specific message exchanges.
The main contribution of our work is showing that (a) it is possible to learn receiver preferences from their message exchanges and stated features; and (b) applying the learned probabilistic model of user preferences in our two-sided market formulation increases the chances of successful matches.

To test our approach we use three months of recorded messages exchanges and user profiles of 2 million distinct male and female pairs of users at Baihe, a large Chinese dating website.
Our results show that the two-side market formulation together with the learning of user preferences increases in up to 48\% the rate of successful matches (as measured by the rate of first contact replies) with respect to recommendations based on the suitor's preference alone. 
We also argue that graph-based recommendation systems are not ideal for large sparse contact graphs such as the one observed at Baihe.

The outline of this work is as follows. 
Section~\ref{sec:background} presents the modeling of the two-side matching market. Section~\ref{sec:learningLDA} introduces an LDA model to learn user preferences.
Section~\ref{sec:exp} describes our experiments. Finally, sections~\ref{sec:related} and~\ref{sec:conclusions} present the related work and conclusions, respectively.


\section{Two-sided Matching Market}
\label{sec:background}
Balancing the expectations of the initiator and the receiver is a challenging task.
This balance is achieved when the website operator cleverly enforces that a recommendation occurs only if both the initiator and receiver would be interested in the match.
To provide a solid theoretic footing to the above idea and, most importantly,  to motivate the importance of learning the receiver preference, we formulate the matching problem as a two-sided matching market.

The two sides of the market refer to the two types of agents in the system (males and females) and a match is the recommendation of a male to a female or vice-versa.
Note that unlike the original formulation of matching markets (such as Gale and Shapley's formulation~\cite{gale1962college}, see Roth and M. Sotomayor~\cite{Roth:1992ws} for a review of two-sided market problems), we allow multiple ``matches'' by allowing multiple recommendations to the suitor and the same receiver be recommended to multiple suitors.
However, we enforce a cap in the average number of (unread) messages a receiver gets per day, which ultimately determines the number of times the receiver can be recommended.

The website wants to provide recommendations that -- under the constraint that no receiver will be inundated with messages (flow control) -- either
 maximize the total number reciprocated messages ({\em max utility}), or any attempt to make a recommendation that increases the reply rate of any participant necessarily results in the decrease in the reply rate of some other participant with an equal or smaller reply rate ({\em max-min fair}).
In what follows we present the max utility optimization problem. 
Extending the optimization to max-min fairness is trivial.

Formally, let $V$ denote the set of website users.
The indicator function that tells if two users $s,r \in V$  are on opposite sides of the maker is
\[
\delta_{sr} = \begin{cases}
                        1 & \text{if $s$ and $r$ are in opposite sides of the market,} \\
                        0 & \text{otherwise.}
                    \end{cases}
\]
Let $x_{sr}$ be the probability that user $s$ is recommended to user $r$. 
If $s$ and $r$ are on the same side of the market, i.e., $\delta_{sr} = 0$, then $x_{sr} = 0$, otherwise $0 \le x_{sr} \le 1$.
%
The following functions define the two-sided market optimization:
\begin{itemize}
\item $f(s,r)$ is the probability that $s$ initiates communication upon receiving a recommendation of user $r$.
\item $g(r,s)$ is the probability that $r$ replies to $s$.
\item $C_{S}(s)$ is the expected maximum number of messages that user $s$ can send
during a day (suitor capacity), $r\in R$
\item $C_{R}(r)$ is the expected maximum number of messages that user $r$ should
receive during a day (receiver capacity)
\end{itemize}

Most of the focus of this work is on learning $f$ and $g$ and showing that there is much to gain when considering receiver preferences.
The values of $C_S$ and $C_R$ are determined by the website operator.
Using the above definitions the max expected utility optimization is then
\begin{equation}\label{eq:opt}
\max\sum_{s\in V}\sum_{r \in V}  f(s,r)g(r,s)x_{sr} \delta_{sr}  \: ,
\end{equation}
subject to
\begin{align*}
\sum_{\forall s\neq r} g(r,s) f(s,r) x_{sr} \delta_{sr}  &  \le C_{R}(r),\:\forall r ,\\
\sum_{\forall r\neq s}  g(r,s) f(s,r) x_{sr} \delta_{sr}  & \le C_{S}(s),\:\forall s,\\
x_{sr} & \in (0,1),\:\forall s,r.
\end{align*}
The above optimization problem can be easily solved with any off-the-shelf linear program package.
An online fully distributed solution, however, requires introducing the dual and using shadow pricing to coordinate~\cite{kelly1998rate}
the recommendations across different servers, as task that is part of our future work.
In what follows we focus on our main goal, the more challenging task 
of learning suitor and receiver dating preferences from the data.

It is important to note that $f$ and $g$ are distinct functions; that is, a suitor may avoid contacting users with a given ``undesirable'' trait but, paradoxically, pay little heed to the same trait when  acting as a receiver (Slater~\cite{Slater13Love} showcases a variety of anecdotal examples of such behavior along with the related social science literature that documents this discrepancy).
However, due to the limited amount data of our dataset used to train our learning algorithm (more details about our experiments in Section~\ref{sec:exp}), we observe that 
treating $f$ and $g$ separately has an adverse effect in the number of samples used to train our model and thus our ability to correctly learn
the true user preferences.
Hence, in what follows we assume that $f$ and $g$ are equivalent ($f \equiv g$)
in order to use all message exchanges regardless to whether the user acts as a suitor or as a receiver.

\section{Learning Dating Preferences}\label{sec:learningLDA}
In this section, we first define user representation, user type and user preference for the online dating network. We introduce the Latent Dirichlet Allocation (LDA) model and modify it to learn user revealed preferences. 
\subsection{Dating Dataset}
Our data consists of 200,000 uniformly sampled newly registered users in the month of November, 2011 from Baihe.com's Chinese dating website. 
It includes 139,482 males and 60,518 females, with each gender making up 69.7\% and 30.3\% of the sampled users, respectively. 
Users come from all over China and also abroad~\cite{XRCLT13}.
For each user we obtain all incoming and outgoing messages from the date that the account was created until January 31st, 2012.
We also obtain profile information of all parties involved in these message exchanges, totaling 2 million unique pairs of users exchanging messages during our observation period.  
The content of each message is removed for privacy concerns but other relevant information remains, such as the message timestamp, the suitor's and receiver's profiles, which consists of 21 features including gender, age, registration timestamp, blood type, weight, height, education, occupation, annual income level, housing situation (renting, home owner), body type, Western zodiac sign, Chinese zodiac sign, number of profile photos,  whether user owns a car, city of residence, and whether users has a child and lives with the child, among other characteristics.


\subsection{Selection of Relevant Features}
In building a probabilistic model of user preferences, we first significantly reduce the  problem dimension by eliminating features that have little predictive power on the likelihood that a user will send or reply a message.
Before we reduce the number of features between pairs of users, we first expand the feature set to also include differences in age, height, weight, education, and income, and whether or not the pair has the same marriage and housing status. 

To model user preference, we seek features that are strongly correlated with the {\em reply} feature, as a {\em reply} indicates user interest. We use two techniques to measure the corelation between {\em reply} and other features: the score of {\em information gain ratio}~\cite{dietterich2000experimental,kohavi1996scaling} and ``variable importance score'' from random forests~\cite{archer2008empirical}. We only keep variables with both scores higher than average and removed the rest.

After that, there could be still variables containing the same information to decide ``reply'' feature.  For example, {\em age} and {\em Chinese zodiac sign},  may be highly correlated and thus we only need to include one of them, as the feature {\em Chinese zodiac sign} has 12 values representing the year when the user is born. We measured the ``information similarity'' between two variables with the conditional entropy and the mutual information of each pair of features. Note that a small conditional entropy means that the feature is largely determined by the other. A large mutual information means two features share information. A feature will be eliminated if there exists another feature that contains most of its information about the {\em reply} value. 
For instance, using the above age and zodiac example we observe that the conditional entropy of {\em Chinese zodiac sign} given {\em age} is {\em H}(Chinese zodiac\textbar age$)\approx 0$. 

We identify as the five most relevant features: {\em age, weight, income difference, children information and height difference}. 
Throughout the reminder of the paper we refer to this five-feature tuple $v=$(age, weight, incomeDif,  childInfo, heightDif) as the feature vector of a user. 
The large number of unique values of {\em age}, {\em weight}, and {\em height} complicates our information gain analysis. 
To ameliorate this problem we  apply the ChiMerge algorithm, a bottom-up Chi-square quantization algorithm~\cite{liu1995chi2}.  
%
After discretization, feature {\em age} has seven intervals, {\em weight} nine, {\em height} 11, making 21 intervals in {\em height difference}. 
For each gender, we define the set of all possible feature tuples $V=\{v_d\}_{d=1}^{D}$. 



\subsection{Latent Dirichlet Allocation (LDA) to\\ Uncover Latent User Preferences}\label{sec:LDA}
Now that the set of relevant features is defined, we turn our attention to grouping users into $T$ ($T$ is a constant) user types according to their latent dating preferences.
To simplify our notation without loss of generality in what follows we consider the suitors to be all on the same side of the maker (say, females) and the receivers all to be on the other side of the market (say, males).
Latent Dirichlet Allocation (LDA) is a powerful statistical technique widely used in Topic Modeling in Natural Language Processing~\cite{blei2003latent}. 
LDA defines a group of latent variables and, through Bayesian inference, reveals the relations between latent topics and the observed documents. 
These learned latent topics determine the similarity between documents and can be used to classify them.

\begin{figure}[ht]
\centering
\includegraphics[scale=0.3]{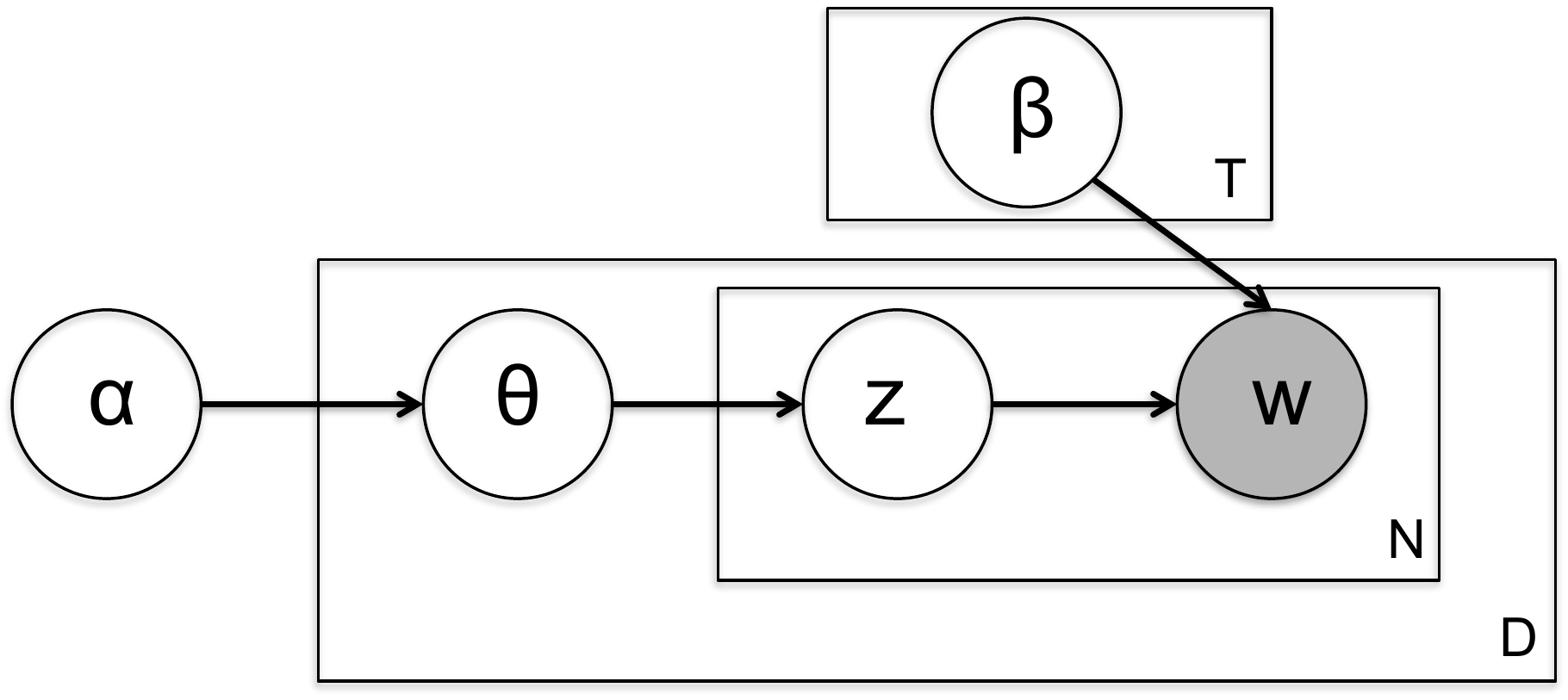}
\label{fig:LDAconversion}
\caption{LDA graphical model of user preference.}
\end{figure}

Similarly, our model makes use of the observed message exchanges to learn user dating preferences. 
Figure~\ref{fig:LDAconversion} shows our graphical model.
Users have latent ``types'' that follow distribution $\vec{\theta}=(\theta_1,\cdots,\theta_T)$. The value of $\vec{\theta}$ is drawn from a Dirichlet distribution $Dir(\vec{\theta};\alpha \vec{m})$ $= \frac{\Gamma(\alpha)}{\prod_{t=1}^{T}\Gamma(\alpha m_t)}\prod_{t=1}^{T}\theta_t^{\alpha m_t-1}$ with $\alpha>0$ and $\sum_i m_i = 1$. 
Let $D$ denote the number of users that send (initiate or reply) at least one message in the training data
and $N$ the total number of such messages.
Let $\vec{z} = (z_d)_{d=1}^D$ denote the user types drawn i.i.d.\ from the distribution $\vec{\theta}$.
User $d$ contacts (i.e., either initiates messages or replies to received messages) $k_d > 0$ users whose feature sets are defined as $\vec{w}_d = (w_{1,d}, w_{2,d},\cdots,w_{k_d,d})$.

It is crucial to determine how user $d$ chooses to engage in message exchanges with other users on the other side of the market.
In our model the probability that user $d$ contacts a set of $k_d$ users with feature values $\vec{w}_d$ is $P(w_{1,d}, w_{2,d},\cdots,w_{k_d,d}|t)  =  P(w_{1,d}|t) \cdots P(w_{k_d,d}|t) = \prod_{i=1}^{k_d}\phi_{w_{i,d}|t}$, where $\phi_{w_{i,d}|t}$ is a parameter in categorical distribution $\vec{\phi_t}=(\phi_{v|t})_{v\in V}$. LDA model assumes the values of $\vec{\phi_t}$ follows a Dirichlet distribution $Dir(\vec{\phi_t};\beta\vec{n})= \frac{\Gamma(\beta)}{\prod_{v\in V}\Gamma(\beta n_v)}\prod_{v\in V}\phi_{v|t}^{\beta n_v-1}$ with  hyperparameters $\beta$ and $\sum_i n_i = 1$. 

\noindent
{\em Likelihood functions.}
 The probability that the model generates the observed message exchanges in the data, observations formally defined as $Data = (\vec{w}_1,\ldots,\vec{w}_D)$, is 
 \begin{align}
 P(Data| \vec{z},\Phi,\vec{\theta})   &=	\prod_{d=1}^{D}\prod_{i=1}^{k_d}P(w_{i,d}|z_d,\Phi)	\nonumber\\
 	 			     & = \prod_{t=1}^{T} \prod_{v\in V}\phi_{v|t} ^{N_{v|t}},
  \label{eq:likelyhood2}
 \end{align}
%
%
where  $\Phi=\{\vec{\phi_t}\}_{t=1}^T$. The posterior distribution is obtained using Bayes rule
\begin{align}
&P(\Phi|Data) = \frac{P(Data|\Phi,\theta,\vec{z})P(\Phi)P(\vec{z}|\theta)P(\theta)}{P(data,\vec{z})p(\theta)} \nonumber \\
	&= \prod_{t=1}^{T}Dir\left(\vec{\phi_t};\left(\frac{\beta  n_1+N_{1|t}}{\beta+ N_t},\cdots,\frac{\beta n_{|V|}+N_{|V||t}}{\beta+ N_t}\right)\right). \label{eq:Phiposterior}
\end{align}
where $N_{i|t}, (i=1,\cdots,|V|)$ is the number of messages from type $t$ suitor to receiver with feature tuple $v_i$, $\sum_iN_{i|t}=N_t$.

%
%
%
%
Similarly, the type of user $d$ given evidence $Data_{(-d)}$, where  $Data_{(-d)}$ denotes $Data$ without user $d$ messages, is  
\begin{align}
P(z_{d}=t| Data_{(-d)}) &= \int_{\theta} P(z_{d}=t|\theta,Data_{(-d)})P(\theta|Data_{(-d)}) d\theta \nonumber \\
& =\frac{D_t+\alpha m_t}{D+\alpha},
\label{eq:zD}
\end{align}
where $D_t$ is the number of users of type $t$ and $D = \sum_{t=1}^T D_t$.

\noindent
{\em Learning user preferences through Gibbs sampling.}
Estimating $\Phi$, $\vec{\theta}$, and $\vec{z}$ from the data through maximum likelihood requires a combinatorial number of iterations.
Hence, we resort to Gibbs sampling to estimate the model parameters from the data. 
Each user $d$ with user type $z_d$ sends messages to a set of receivers $W_d = \{w_{i,d}\}$. 
Let subscript ${(-d)}$ denote a data structure without user's $d$ variable. 

Using Gibbs sampling we sample the value of $z_d$ given $\vec{z}_{(-d)}$ and $Data_{(-d)}$ with probability 
\begin{align*}
P(z_d|Data,\vec{z}_{(-d)})&=\frac{P(\vec{w}_d,z_d|Data_{(-d)},\vec{z}_{(-d)})}{\sum_{z_d}P(\vec{w}_d,z_d|Data_{(-d)},\vec{z}_{(-d)})} \\
& \propto P(\vec{w}_d,z_d|Data_{(-d)},\vec{z}_{(-d)}),
\end{align*}
and substituting Eqs.~\eqref{eq:Phiposterior} and~\eqref{eq:zD}  into the above expression  yields
\begin{align*}
&P(\vec{w}_d,z_d=t|Data_{(-d)},\vec{z}_{(-d)}) = \frac{P(Data,\vec{z})}{P(Data_{(-d)},\vec{z}_{(-d)})} \nonumber\\
	&\propto \frac{\Gamma(N_t^{(-d)}+\beta)}{\prod_{v}\Gamma(N_{v|t}^{(-d)}+\beta n_v)} 
			\frac{\prod_{v}\Gamma(N_{v|t}+\beta n_v)}{\Gamma(N_t+\beta)}
			  \frac{D_t^{(-d)}+\alpha m_t}{D-1+\alpha}  \nonumber ,
\end{align*}
where $N_{v|t}^{(-d)}$ is the number of receivers with the $v$-th feature tuple receiving from type $t$ user in $Data_{(-d)}$, $N_t^{(-d)}=\sum_{v}N^{(-d)}_{v|t}$, and $N_t=\sum_{v} N_{v|t}$.

\subsection{Application to Two-sided Markets}
In Section~\ref{sec:background} we introduced the two-side matching market with preference functions $f(s,r)$ and $g(r,s)$.
We then made the simplifying assumption that $f \equiv g$.
In what follows we obtain $f$ (or $g$) from the data using our LDA results.
Let $\mu^{(d)}_t = P(z_{d}=t| Data)$ and  $v_d$ the relevant feature vector of user $d$.
Using the learned user mixture types and preferences we can now define function $f$ and $g$ for the any user pair $(s,r)$:
\begin{align} \label{eq:fg}
f(s,r) & = g(s,r) = \delta_{s,r} \sum_{t=1}^{T} \mu^{(s)}_t\phi_{v_r|t}\,, \quad \forall s,r.
\end{align}

\subsection{Two-sided Markets \& New Users}
%
We use the above LDA model to estimate $P[z_d=t|\vec{w}_d]$, the probability that a user $d$'s  user type $z_d=t$ given his messages.
After that, $\vec{\phi_t}$, the preference of the user type $t$, is assigned to him.
However, we would like to say something about users without observed message exchanges. 
A reasonable way to solve this problem is to use the user profile to predict the user type. 
We assume the relevant features in a user's profile have strong correlation with his user type, in that case, we can use maximum-likelihood estimation (MLE) to obtain the probability of the user type given his features $v_{d}$:  $q^{(d)}_t=P(z_d=t|v_d)$.
For these users we can construct a mixture of preferences from user $s$ to a user $r$ with feature vector $v_r$:
\begin{align}
\hat{f}(s,r) =  \delta_{s,r} \sum_{t=1}^{T}q^{(s)}_t \phi_{v_r|t},
\label{eq:mixPref}
\end{align}
where $\hat{f}(s,r)$ is the probability that user $s$ initiates (or replies) a message to user $r$ that has feature vector $v_r$.
In what follows we use our data in combination with Eq.~\eqref{eq:mixPref} and the two-sided market formulation to significantly improve the success rate of recommended matches.


\section{Results}
\label{sec:exp}
In this section, we first measure how well the LDA model can learn user preferences using synthetic data. 
We then evaluate the gains obtained from recommending Baihe users based on the learned preferences from the Baihe data (with the techniques described in Section~\ref{sec:LDA}) and two-sided market principles introduced in Section~\ref{sec:background}.

\subsection{Results with Synthetic Data}

To verify whether the LDA model can truly learn user preferences we simulate a dating market (since we cannot perform live experiments at Baihe and there is no ground truth in the Baihe dataset). 
We generated 20,000  male and female users with profiles, respectively. Each simulated user has a feature vector (\textit{age, has/lives with children, weight, income, height}). We calculated the marginal distribution of each feature sample them from their empirical distribution in the Baihe data. 

Our simulator uses eight distinct user types, four types per gender. 
For each gender, the user preference of type $t$ is a distribution over all feature vectors, denoted as $p_t= (p_{{v_1}|t}, \ldots, p_{v_{|V|}|t})$, where $v\in V$ is a feature vector and $t=\{(i,q):i=1,\ldots4,q\in\{\text{male},\text{female}\}\}$.
Each user type has, potentially, a different set of favorite feature vectors such that users of that type exchange messages differently than users of other types. 
We then randomly select 5\% of the feature vectors in $V$ that belong to the opposite gender as type $t$'s favorite feature vectors, denoted as $F$. Then for each $v \in F$ we set $p_{v|t}$ with a value drawn uniformly from  interval $(300,500)$. 
For the remaining feature vectors, $v \in V \backslash F$, $p_{v|t}$ is sampled uniformly from the interval $(1,2)$. 
Finally, we normalize $p_{t}$ such that $\sum_{v \in V} p_{v|t}=1$.


To simulate the dating dynamics we randomly recommend 100 users of the opposite gender to each user, henceforth denoting the set of recommended users $L$. 
Each user then chooses $k_d$ receivers among the 100 recommendations, where $k_d$ is a value uniformly sampled from $\{0,\ldots,10\}$.
The $k_d$ lucky receivers are chosen from user set $L$ through a multinomial distribution with parameters $100$, $k_d$, and $(p_{v|t})_{v \in L}$.
For the LDA estimation we set the maximum number of user types $T = 10$ for each gender in order to test the impact of having more user types in the model than the data allows.
%
%
%
%
The goal of this experiment is to test if the LDA model can correctly learn the four preferences for each of the genders.

\begin{table}[ht]
\caption{Matching Male User Type}
\centering
\scriptsize
\begin{tabular}{lccc}
Type & Precision & Recall & K-L divergence\vspace{1pt} \\
\hline \vspace{-8pt}
\\
type 1 & 98.8\% & 99.8\% &  8.578e-05 \\ 
type 2 & 98.2\% & 99.9\% &  -9.013e-05 \\ 
type 3 & 99.3\% & 98.6\% &   7.401e-05 \\ 
type 4 & 100\% & 100\% &   6.515e-05 \\ 
\hline 
\end{tabular} 
\label{tab:MaleType}
\caption{Matching Female User Type}
\begin{tabular}{lccc}
Type & Precision & Recall & K-L divergence\vspace{1pt} \\
\hline \vspace{-8pt}
\\
type 1 & 96.8\% & 99.4\% & -1.463e-04 \\ 
type 2 & 99.7\% & 99.8\% & 9.117e-05 \\ 
type 3 & 98/3\% & 98.4\% & 1.863e-04 \\ 
type 4 & 98.5\% & 96.8\% & -1.421e-04 \\ 
\hline 
\end{tabular} 
\label{tab:FemaleType}
\vspace{-5pt}
\end{table}

Our results show that our model classifies most males  (99.5\%) and females  (99.6\%) into one of four large user type groups, showing that despite the maximum number of user types of each gender being large, $T = 10$, the model is able to learn the correct number of distinct user types (four) for both genders. 
Focusing only on these four largest estimated groups (of user types) of each gender we now compare the true preferences, $p_t=(p_{v_1|t},\ldots, p_{v_{|V|}|t})$, 
against the learned preference from our model, $\phi_t$.
For this comparison we use the K-L divergence between $p_t$ and $\phi_t$:\footnote{\scriptsize To solve the problem of matching the correct learned user type label $t$ with the true user type label we consider a bipartite graph $G(E_g,V_g)$ with nodes of true preferences $p_i\in V_g$ on one side and  nodes of the revealed preferences $\phi_j\in V_g$ on the other, the weight of the edge $e_{i,j}= e(p_i, \phi_j)\in E_g$  is the K-L divergence $D_{KL}(p_i||\phi_j)$. We  can match the defined preferences to the revealed preferences by solving the minimum weight matching in polynomial time. } $$D_{KL}(p_t||\phi_t)= \sum_{v=1}^V\log\left(\frac{p_{v|t}}{\phi_{v|t}}\right) p_{v|t}.$$
%
%
Tables~\ref{tab:MaleType} and~\ref{tab:FemaleType} show the precision and  recall of each estimated user type for males and females, respectively. 
The precision and recall are close to 100\%, showing that the LDA estimation indeed was able to accurately recover the user type with just a few observed messages (in average 4.5 per user) .   
Also note that the K-L divergences are low, suggesting that the estimated and true preferences are remarkably similar. 

\subsection{Baihe Results}
In this section we focus on testing whether the two-sided matching recommendations can improve the number of successful matches in the Baihe dataset.
Henceforth we denote ``probability that a suitor message is replied'' as the {\em success rate}.
Recall that the success rate is the utility function in that we seek to maximize in Eq.~\eqref{eq:opt}.
Our experiment obeys the following principle: we eliminate half of the messages sent from suitors to receivers. 
For each suitor in the dataset that has messages sent to two or more distinct receivers, we use the distinct recommendation algorithms to
 choose which message stay in the dataset and which message are discarded.
We then compare the performance of the recommendation algorithms by contrasting the average success rate of the messages that stayed in the dataset.

In the above experiment we compare three recommendation algorithms: (a) random, (b) suitor preference ($f(s,r)$), and (c) two-sided (suitor and receiver) preference ($f(s,r)g(r,s)$).
We first use LDA model to learn the user preferences in the training set.
We then assign those preferences to the users in the testing set with the mixture model.
First we partition the suitors into ten equal size datasets $\{U_i\}_{i=1}^{10}$ such that there are no messages between the users in distinct partitions.
We use nine randomly selected dataset partitions to train the LDA model and the one partition not used for training is used to test our algorithm; without loss of generality we denote the test partition $U_{10}$. 
This training-test procedure is known as ten-fold cross validation.

We rank the messages sent by each suitor $s \in U_{10}$ to its receivers $\{r_i\}_{i=1}^{k_s}$ according to either $\hat{f}(s,r_i)$ if the recommendation just uses the suitor preference or $\hat{f}(s,r_i)\hat{f}(r_i,s)$ if it is a two-sided recommendation, where $\hat{f}(s,r)$ is as described in Eq.~\eqref{eq:mixPref}.
We must use $\hat{f}$ of Eq.~\eqref{eq:mixPref} instead of $f$ of Eq.~\eqref{eq:fg} as $s$ and $r_i$ are in the test set, i.e., our learning algorithm was not trained with the message exchanges of $s$.
We then keep the top half of the ranked messages and discard the rest of the messages.
Our measure of goodness is the success rate of the top half ranked messages. 

\begin{figure}[ht]
\centering
\includegraphics[height=1.1in,width=2.5in]{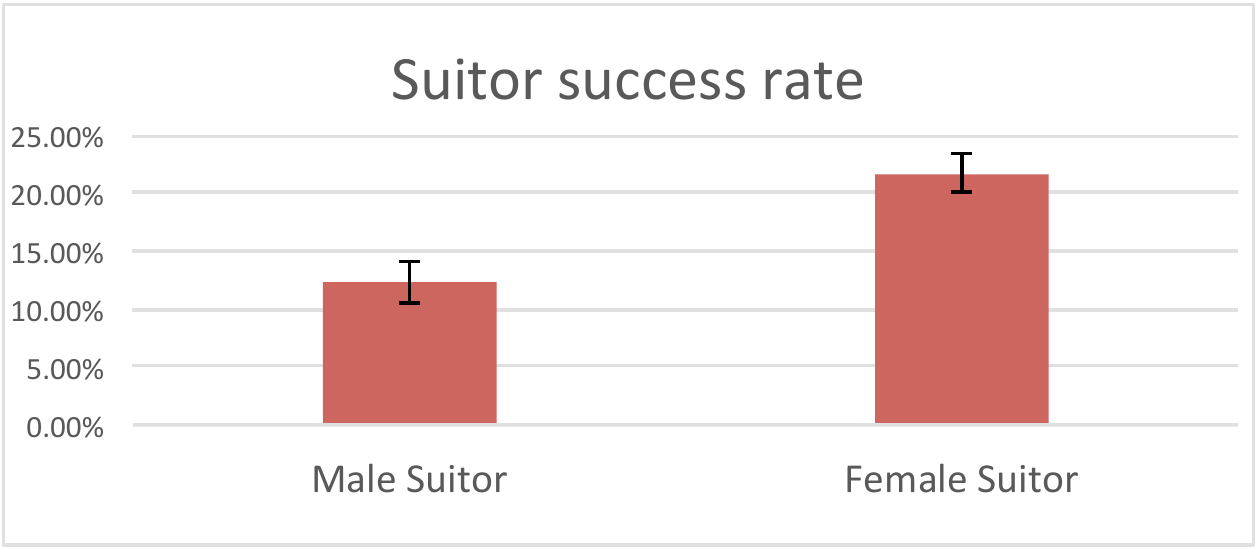}
\caption{Success rate of one-sided suitor-based recommendations.}
\label{fig:joint_rplrate}
\end{figure}

\begin{figure}[ht]
\centering
\includegraphics[height=1.5in,width=2.5in]{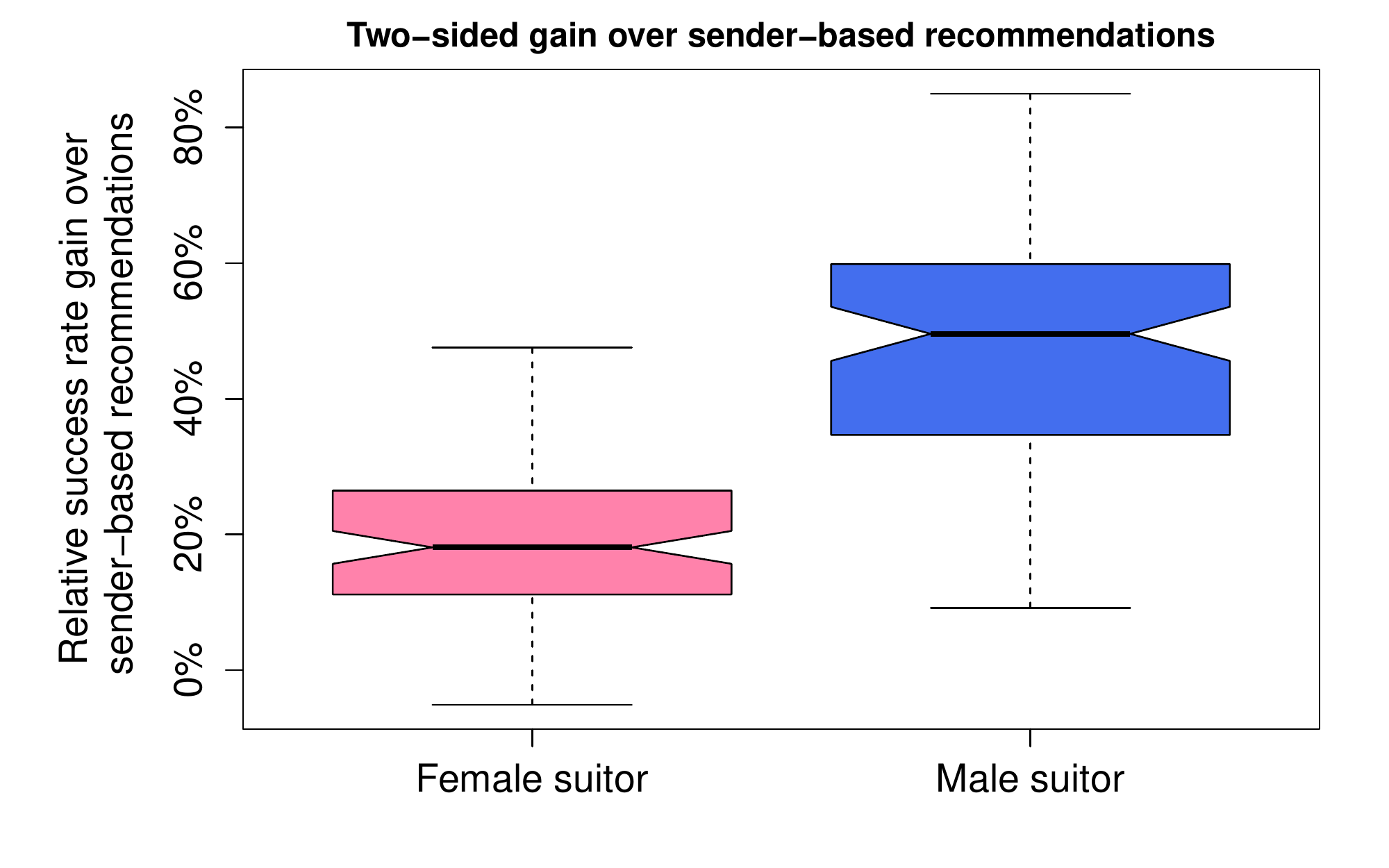}
\vspace{-10pt}
\caption{Relative gain in success rate of two-sided over one-sided suitor-based recommendations.}
\label{fig:jointPref}
\end{figure}

Figure~\ref{fig:joint_rplrate} shows average success rate experienced by male and female suitors based on the suitor preference.
Interestingly, these success rates are the same as in random selection.
Male suitors have a much lower success rate, with an average 12.2\% chance of having their messages replied, while females are significantly more successful, with an average of 21.7\% success rate.
The black bars in Figure~\ref{fig:joint_rplrate} shows the standard deviation of our experiments.
We now contrast the above results with the success rate of messages selected based on two-sided preferences.
Figure~\ref{fig:jointPref} shows a box plot of the relative percentage gain of success rate of two-sided preferences over the success rate using suitor preferences alone.
Male suitors have a significant improvement in their success rate showing an median of 46.84\% higher success rates.
Female suitors also show a median improvement of 16.5\% higher success rates.
These experiments indicate that  two-sided framework can achieve more successful matchings than traditional suitor-only recommendations.
 

\begin{figure}[ht]
\centering
\includegraphics[width=2.5in,height=1in]{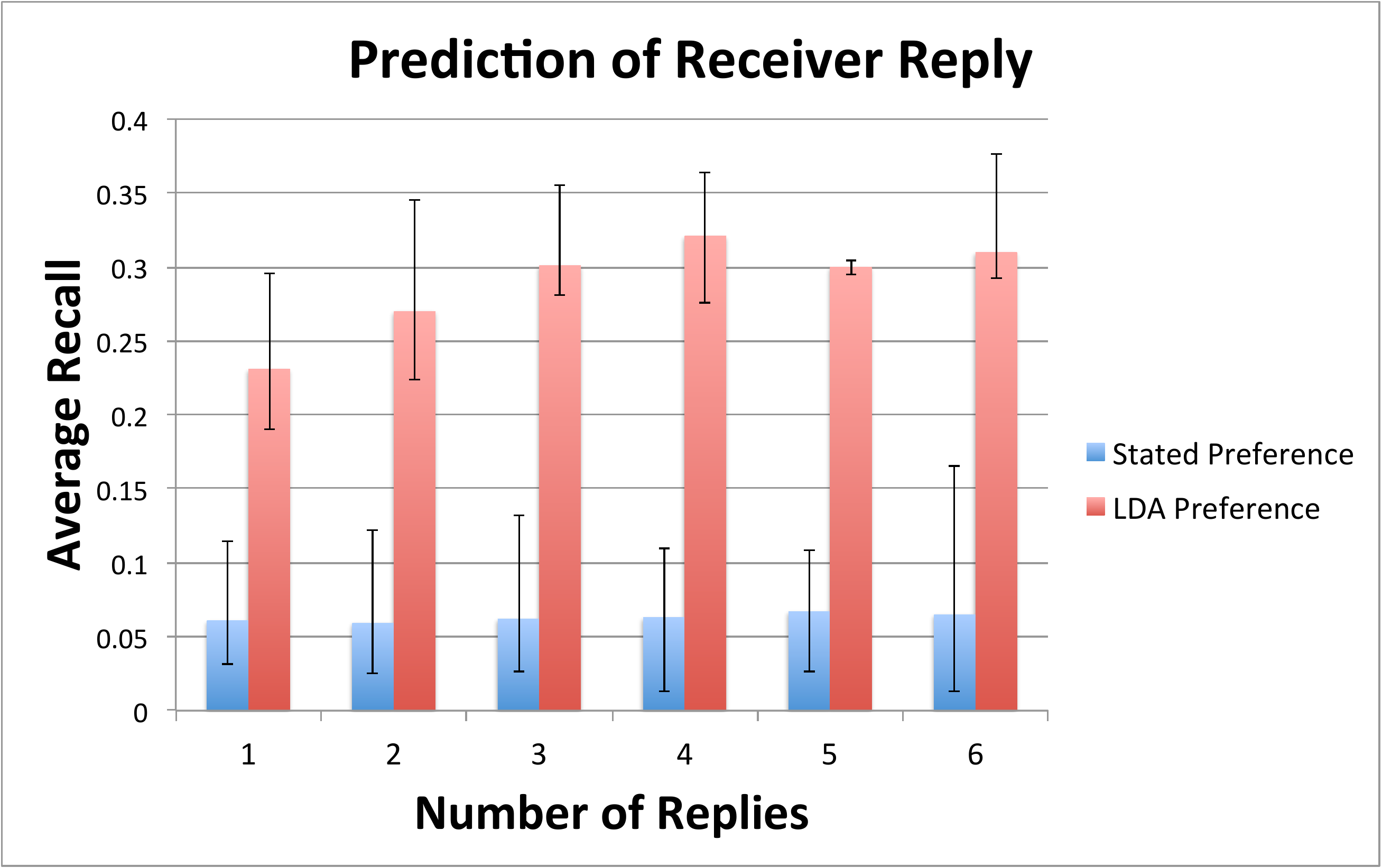}
\caption{LDA learned preferences are better predictors of user reply than their stated preferences.}
\label{fig:recallRec}
\end{figure}

\noindent
{\em LDA preferences v.s.\ stated preferences.}
In Baihe users can state features of their preferred mates.
To test whether LDA preferences are more predictive of the true preference than 
the user's stated preference we test the predictive power of LDA learned preferences against the user stated preferences.
Figure~\ref{fig:recallRec} shows the probability of a receiver reply given his or her LDA and stated preferences. 
The LDA learned preferences of the receivers clearly outperform their stated preferences. 

\section{Related Work}\label{sec:related}
Online match-making ``user-to-user'' recommendation systems differ from ordinary ``user-to-item'' in that a match is only successful if both sides (suitor and receiver) agree that the match is good~\cite{brozovsky2007recommender,kutty2013people,zweig2011systematic}. 
Recently there has been much effort in building recommendation systems based on the ``user-to-user'' matching concept~\cite{Adachi:2003ic,Alsaleh:2011hf,brozovsky2007recommender,cai2010learning,Chen:kg,Diaz:2010hf,Hitsch:2010wm,krzywicki2010interaction,Nayak:hm,Pizzato:2010jg}.

The majority of the ``user-to-user'' online dating recommendation systems are graph-based collaborative filtering algorithms~\cite{brozovsky2007recommender,cai2010learning,Chen:kg,kutty2013people}.
For instance, Kutty et al.~\cite{kutty2013people} proposed a graph mining technique that calculates the similarity of the users' preferences and the similarity of user profiles according to both the users' stated preferences and the structure of ``user-attribute bipartite network''. 
Unlike online social networks (OSNs), where collaborative graph-based filtering makes sense due to the highly clustered nature of OSNs, the bipartite matching graph tends to be highly sparse.
For a graph-based collaborative filter to work as a recommendation system, the recommended matchings must be artificially clustered by recommending the same set of ``receivers'' to suitors that are deemed similar.
This approach, however, creates the odd situation where similar suitors are artificially forced to compete for the same set of possible dates.

A more theoretically sound approach to two-sided matching markets is found in the work of Adachi~\cite{Adachi:2003ic}. 
Adachi introduces a search cost penalty to the Gale-Shapley two-sided matching formulation~\cite{gale1962college}. 
Hitsch et al.~\cite{Hitsch:2010wm} uses Adachi's formulation, together with an interesting psychological study of the matching market and user preferences, to argue that Adachi's algorithm can be combined with a feature-based logistic regression as a recommendation system for online dating.
Our framework has significant of advantages over that of Adachi~\cite{Adachi:2003ic} and Hitsch et al.~\cite{Hitsch:2010wm}.
First, our probabilistic framework (Eq.~\eqref{eq:opt}) avoids the unnecessary computational hardness and sub-optimality of binary optimization problems.
Second, in our framework preferences are seen as probabilities, making it easier to  map the output of probabilistic models (e.g., LDA) to the implementation of the algorithm.
Third, unlike  Adachi's framework, there is no abstract ``search cost penalty''. 
Rather, our framework constraints are intuitive to website operators: the average number of recommendations to a single user and the average number messages a user should receive. 
Finally, unlike the feature-based logistic regression used in  Hitsch et al.~\cite{Hitsch:2010wm},
we propose a model that is also able to tailor the recommendations to the observed user behavior rather than being solely restricted to user features.

\section{Conclusions} \label{sec:conclusions}
In this work we propose a probabilistic two-side matching market framework for online dating recommendations. 
We show that considering preferences of both sides of the market can dramatically improve the number of successful matches. 
We also show how an LDA-based algorithm that learns user preferences can be incorporated into our framework.
In a synthetic dating market we show that our LDA model can successfully classify similar users and learn their preferences. 
Interestingly, by using LDA we gain the ability of using unstructured text (such as user self-descriptions) as features for free.
Our principled  probabilistic two-sided matching framework sheds light into key fundamental principles of online dating matchings.

Our recommendation system is, however, incomplete.
While we believe our framework is both practical and scalable, it has not been implemented in a large live system.
Moreover, a principled approach to incorporate user queries~\cite{Diaz:2010hf} into our framework remains an open problem. 
Replacing LDA with psychological principled models of user preference and behavior may also prove advantageous in our framework, but whether or not other models of user preference can improve upon our simple LDA model remains to be seen.




%
\scriptsize
\bibliographystyle{abbrv}
\bibliography{sigproc}  
%
%

\end{document}